\documentclass{elsart1p}
\usepackage{amsmath}
\usepackage{graphicx}
\usepackage{dcolumn}
\usepackage{bm}
\usepackage[square,comma,sort&compress]{natbib}

\newcommand{\fo}{\ensuremath{^{18}\mathrm{F}}}
\newcommand{\fn}{\ensuremath{^{19}\mathrm{F}}}
\newcommand{\nen}{\ensuremath{^{19}\mathrm{Ne}}}
\newcommand{\pg}{$^{18}$F(p,$\gamma)^{19}$Ne}
\newcommand{\pa}{$^{18}$F(p,$\alpha)^{15}$O}
\newcommand{\dpt}{d($^{18}$F,p)$^{19}$F}
\newcommand{\dpa}{d($^{18}$F,p$\alpha)^{15}$N}
\def\power#1{\mbox{$\times10^{#1}\ $}}
\newcommand{\gap}{\mathrel{ \rlap{\raise.5ex\hbox{$>$}}
                    {\lower.5ex\hbox{$\sim$}}  } }
\newcommand{\lap}{\mathrel{ \rlap{\raise.5ex\hbox{$<$}}
                    {\lower.5ex\hbox{$\sim$}}  } }

\begin{document}
\runauthor{N.~de~S\'er\'eville}
\begin{frontmatter}
\title{Indirect study of $^{19}$Ne states near the $^{18}$F+p threshold.} 

\author[CSNSM]{N.~de~S\'er\'eville\thanksref{A}}
        \corauth[nds]{Corresponding author.}
        \ead{deserevi@fynu.ucl.ac.be}
\author[CSNSM]{A.~Coc}
\author[CRC]{C.~Angulo}
\author[CSNSM]{M.~Assun\c{c}\~ao}
\author[IPNO]{D.~Beaumel}
\author[DAPNIA]{E.~Berthoumieux}
\author[USTHB]{B.~Bouzid}
\author[CRC]{S.~Cherubini\thanksref{B}}
\author[CRC]{M.~Couder\thanksref{C}}
\author[CRC]{P.~Demaret}
\author[GANIL]{F.~de~Oliveira Santos}
\author[LNS]{P.~Figuera}
\author[IPNO]{S.~Fortier}
\author[CRC]{M.~Gaelens\thanksref{D}}
\author[GSI]{F.~Hammache\thanksref{E}}
\author[CSNSM]{J.~Kiener}
\author[CSNSM]{A.~Lefebvre-Schuhl}
\author[IMRE]{D.~Labar}
\author[CRC]{P.~Leleux}
\author[CRC]{M.~Loiselet}
\author[CRC]{A.~Ninane}
\author[USTHB]{S.~Ouichaoui}
\author[CRC]{G.~Ryckewaert}
\author[SRPUG]{N.~Smirnova}
\author[CSNSM]{V.~Tatischeff}
\thanks[A]{Present address: Universit\'e catholique de Louvain, 
	   Louvain-la-Neuve, Belgium.}
\thanks[B]{Present address: Ruhr-Universit\"at-Bochum, Bochum, Germany.}
\thanks[C]{Present address: Nuclear Physics Department, University of Notre
	   Dame, Indiana, USA.}
\thanks[D]{Present address: International Brachytherapy, Seneffe, Belgium.}
\thanks[E]{Permanent address: IPN, Orsay, France.}

\address[CSNSM]{CSNSM, IN2P3/CNRS and Universit\'e Paris Sud, 
		F-91405 Orsay campus, France}
\address[CRC]{Centre de Recherches du Cyclotron and Institut de Physique
	      Nucl\'eaire, Universit\'e catholique de Louvain, B-1348
	      Louvain-la-Neuve, Belgium}
\address[IPNO]{IPN, IN2P3/CNRS, and Universit\'e Paris Sud, F-91406 Orsay
               Cedex, France}
\address[DAPNIA]{CEA, DAPNIA/SPhN, F-91191 Gif/Yvette Cedex, France}
\address[USTHB]{USTHB, B.P. 32, El-Alia, Bab Ezzouar, Algiers, Algeria}
\address[GANIL]{GANIL B.P. 5027, 14021 Caen Cedex, France}
\address[LNS]{Laboratori Nazionali del Sud, INFN, Via S.~Sofia, 44 - 
              95123 Catania, Italy}
\address[GSI]{GSI mbH, Planckstr. 1, D-64291 Darmstadt, Germany}
\address[IMRE]{Unit\'e d'Imagerie Mol\'eculaire et Radioth\'erapie 
               Exp\'erimentale, Universit\'e catholique de Louvain, B-1348
               Louvain-la-Neuve, Belgium}
\address[SRPUG]{Department of Subatomic and Radiation Physics, University of
                Ghent, Proeftuinstraat 86, B-9000 Ghent, Belgium}

\begin{abstract}
The early $E \leq$ 511~keV gamma--ray emission from novae depends critically
on the \pa\ reaction. Unfortunately the reaction rate of the \pa\ reaction
is still largely uncertain due to the unknown strengths of low--lying proton
resonances near the \fo+p threshold which play an important role in the nova
temperature regime. We report here our last results concerning the study of
the d($^{18}$F,p)$^{19}$F($\alpha)^{15}$N transfer reaction. We show in 
particular that these two low--lying resonances cannot be neglected. These 
results are then used to perform a careful study of the remaining 
uncertainties associated to the \pa\ and \pg\ reaction rates.
\end{abstract}

\end{frontmatter}

\section{Introduction}
\begin{figure}[h]
  \centering
  \includegraphics[width=8.5cm]{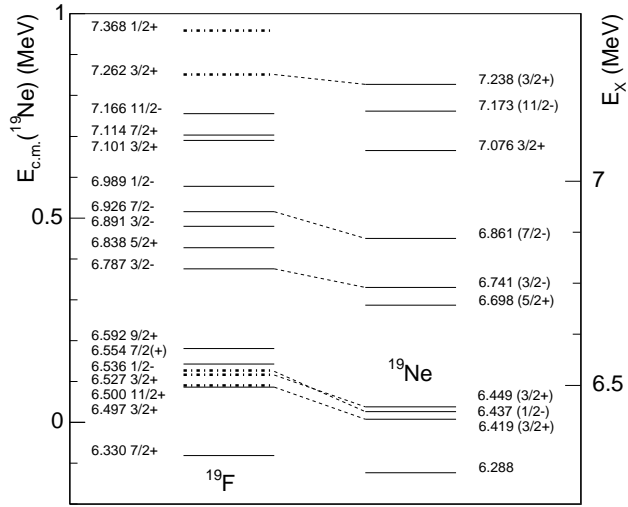}
  \caption{Level scheme for \fn\ and \nen\ above the \fo\ + p threshold.
  Dotted lines connect analog states from~\protect\cite{Utk98}. Dashed-dotted
  lines in \fn\ represent the levels studied in the present work.}
  \label{f:Level}
\end{figure}

Gamma--ray emission from classical novae is dominated, during the first hours,
by positron annihilation resulting from the beta decay of radioactive nuclei.
The main contribution comes from the decay of \fo\ (half--life of 110~min) and
hence is directly related to \fo\ formation during the outburst~\cite{Gom98,
Her99,CHJT00}. A good knowledge of the nuclear reaction rates of production
and destruction of \fo\ is required to calculate the amount of \fo\ synthesized
in novae and the resulting gamma--ray emission. The relevant reactions for the
\fo\ production [$^{17}$O(p,$\gamma$)\fo\ and $^{17}$O(p,$\alpha$)$^{14}$N]
have been recently studied~\cite{Fox04,Fox05,Cha05} and point to a lower
production of \fo. The relevant rates for the \fo\ destruction are the \pg\
reaction ($Q = 6.411$~MeV) and the much faster \pa\ reaction ($Q = 2.822$~MeV).
In order to calculate these
destruction rates, it is important to determine the properties of the \nen\
states near the \fo\ + p threshold. Therefore, an extensive series of
experiments like (p,p) elastic scattering~\cite{Bar00,Gra01,Bar01a}, direct
(p,$\alpha$) measurements~\cite{Coz95,Reh95,Gra97,Gra01,Bar01a,Bar02} and
indirect transfer reactions to populate levels in \nen\ and \fn~\cite{Utk98,
Ser03a,Vis04,Koz05} have been performed. However, its rate remains poorly
known at nova temperatures (lower than 3.5\power{8}~K) due to the scarcity of
spectroscopic information for levels near the proton threshold in the compound
nucleus \nen. This uncertainty is directly related to the unknown
properties of the first three levels in \nen: $E_x$, $J^\pi$ =
6.419~MeV, (3/2$^+$); 6.437~MeV, (1/2$^-$) and 6.449~MeV, (3/2$^+$) (see
Fig~\ref{f:Level}) and following Utku et al.~\cite{Utk98} we will assume the
previous spin and parity assignments even though they have not been measured
directly. The tails of the corresponding resonances at $E_r$ = 8,
26 and 38~keV, respectively, can dominate the \pa\ astrophysical S--factor
in the relevant energy range~\cite{CHJT00}. As a consequence of these nuclear
uncertainties, the \fo\ production in nova and the early gamma--ray emission
was estimated to be uncertain by a factor of $\approx$300~\cite{CHJT00}.

In order to estimate the level position and spectroscopic factors of the
relevant states in \nen, we performed shell--model calculations with the
OXBASH code~\cite{OXBASH}. It was not possible to reproduce experimental
levels at energies about 6--7~MeV, motivating hence an experimental approach.
Due to the very low Coulomb barrier penetrability for the relevant resonances
in \nen, a direct measurement of their strength is at present impossible.
Hence, we have used an indirect method, the \dpt\ transfer reaction, aiming
at the determination of the one--nucleon spectroscopic factor ($S$) in the
analog levels
of the \fn\ mirror nucleus. Assuming the equality of the spectroscopic factors
in analog levels, it is possible to calculate the \nen\ proton widths through
the relation $\Gamma_p = S \times\Gamma_{\mathrm s.p.}$~\cite{Lan60, Mac60}
where $\Gamma_{\mathrm s.p.}$ is the single particle width. Preliminary
results were presented in~\cite{Ser03a}, here we provide a detailed data
analysis.

This paper is organized as follows. The experimental method is detailed in
Sections~\ref{sec:Exp} and~\ref{sec:Spec}. The data analysis is presented
in Sections~\ref{sec:DWBA} and~\ref{sec:Results} and the remaining 
uncertainties concerning the \pa\ and \pg\ reaction rates are studied in the 
concluding Section~\ref{sec:Sfac}.

\section{\label{sec:Exp}Experimental method}
\subsection{Experimental set--up}
The experiment was performed at the CYCLONE RIB facility at the {\it
Centre de Recher\-ches du Cyclotron}, UCL, Louvain--la--Neuve, Belgium.
We used a 14-MeV \fo\ radioactive beam which was produced via the
$^{18}$O(p,n)\fo\ reaction, chemically extracted to form CH$_3$\fo\
molecules, transferred to an ECR source, ionized to the 2$^+$ state and
then accelerated to the relevant energy~\cite{Cog99}. The production of \fo\
was made in the batch mode and an average of 2.2$\times10^6$ \fo\ ions
per second on target was delivered for a total of 40~hours representing 15
\fo\ batches. The beam contamination from its stable isobar $^{18}$O
was evaluated 20~hours after the delivery of the last batch
from the counting rate on the target. The observed counting rate was in fact
compatible with the \fo\ radioactive decay and indicated a $^{18}$O / \fo\
ratio less than 10$^{-3}$.
                                                                                
The \fo\ beam bombarded 100~$\mu$g/cm$^2$ deuterated polyethylene
(CD$_2$) targets. These targets were produced by polymerization and
their thickness and homogeneity were checked by measuring the energy
loss of $\alpha$ particles. The target stoechiometry and contamination
were studied at the Orsay ARAMIS accelerator~\cite{ARAMIS} after the \dpt\
experiment. Elastically scattered protons were detected with a silicon
detector at a laboratory angle of 150$^{\circ}$ for an incident proton
energy of 2.8~MeV~\cite{Ami93, Koch69}. It was found that the stoechiometry
was not modified during the experiment as expected from the low \fo\ beam
intensity.
                                                                                
The products of the \dpt\ reaction were detected using the multi--strip
silicon detector LEDA~\cite{Dav00}. Two configurations named LEDA and LAMP
were used. These detectors are composed of 8 and 6 sectors, respectively,
each divided into 16 radial strips. The energy calibration of the 224 strips
was performed with a calibrated 3$\alpha$--source ($^{239}$Pu, $^{241}$Am
and $^{244}$Cm) whereas the time--of--flight calibration was performed with
a precision pulser.

\begin{figure}[h]
  \centering
  \includegraphics[width=8.5cm]{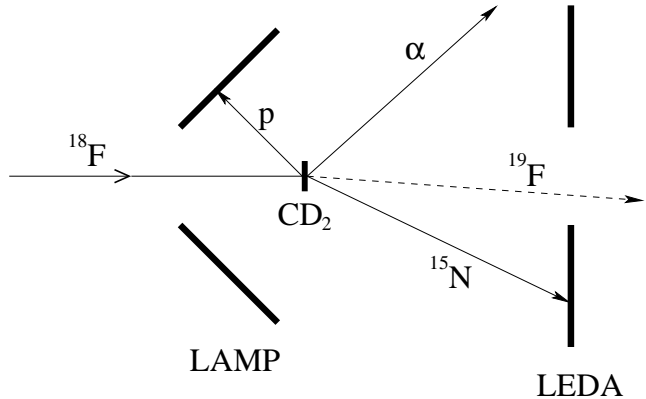}
  \caption{The experimental set--up is shown with \fo\ ions impinging
  on a CD$_2$ target. Protons were detected in LAMP while $^{15}$N or
  $\alpha$--particles from \fn\ decay were detected in coincidence in LEDA.
  See Fig~1 in~\protect\cite{Ser03a} for a perspective view of the set--up.}
  \label{f:Setup}
\end{figure}
                                                                                
The experimental set--up is shown in Fig~\ref{f:Setup}. The protons
from the \dpt\ reaction were detected in LAMP positioned 8.7-cm upstream
from the target. It corresponds to laboratory angles between 110$^\circ$
and 157$^\circ$, i.e. center--of--mass angles between 12$^\circ$ and
44$^\circ$ for the two 3/2$^+$ levels of astrophysical interest.
Since these two levels in \fn\ are situated above the $\alpha$--threshold
at 4.013~MeV, they decay through \fn$^* \to \alpha + ^{15}$N. The decay
products were detected in LEDA positioned 40-cm downstream from the target,
which corresponds to laboratory angles between 7$^\circ$ and 18$^\circ$.
The positions of the detectors were determined by Monte-Carlo simulations:
for LAMP, the angular range and detection efficiency were maximized and for
LEDA, the coincidence efficiency between protons detected in LAMP and
$^{15}$N detected in LEDA was optimized.
                                                                                
Rutherford elastic scattering of \fo\ on the carbon contained in the target,
detected in LEDA, provides the (target thickness) $\times$ (beam intensity)
normalization. We used the strip at 15.7$^\circ$ in LEDA for which the solid
angle is maximum and less sensitive to the exact position of the beam on the
target. Furthermore, for this strip the elastic scattering peaks of \fo\ and
$^{12}$C are well separated. The uncertainty on the normalization is mainly
due to the position of the beam ($\pm$~2~mm) and is estimated to be of 7\%.
                                                                                
Data were collected in event--by--event mode where the multiplicity, the
angle, the deposited energy and the time of flight relative to the cyclotron
radio--frequency were recorded, allowing an off--line analysis of single and
coincidence events.

\subsection{Data reduction}
\begin{figure}[h]
  \centering
  \includegraphics[width=9cm]{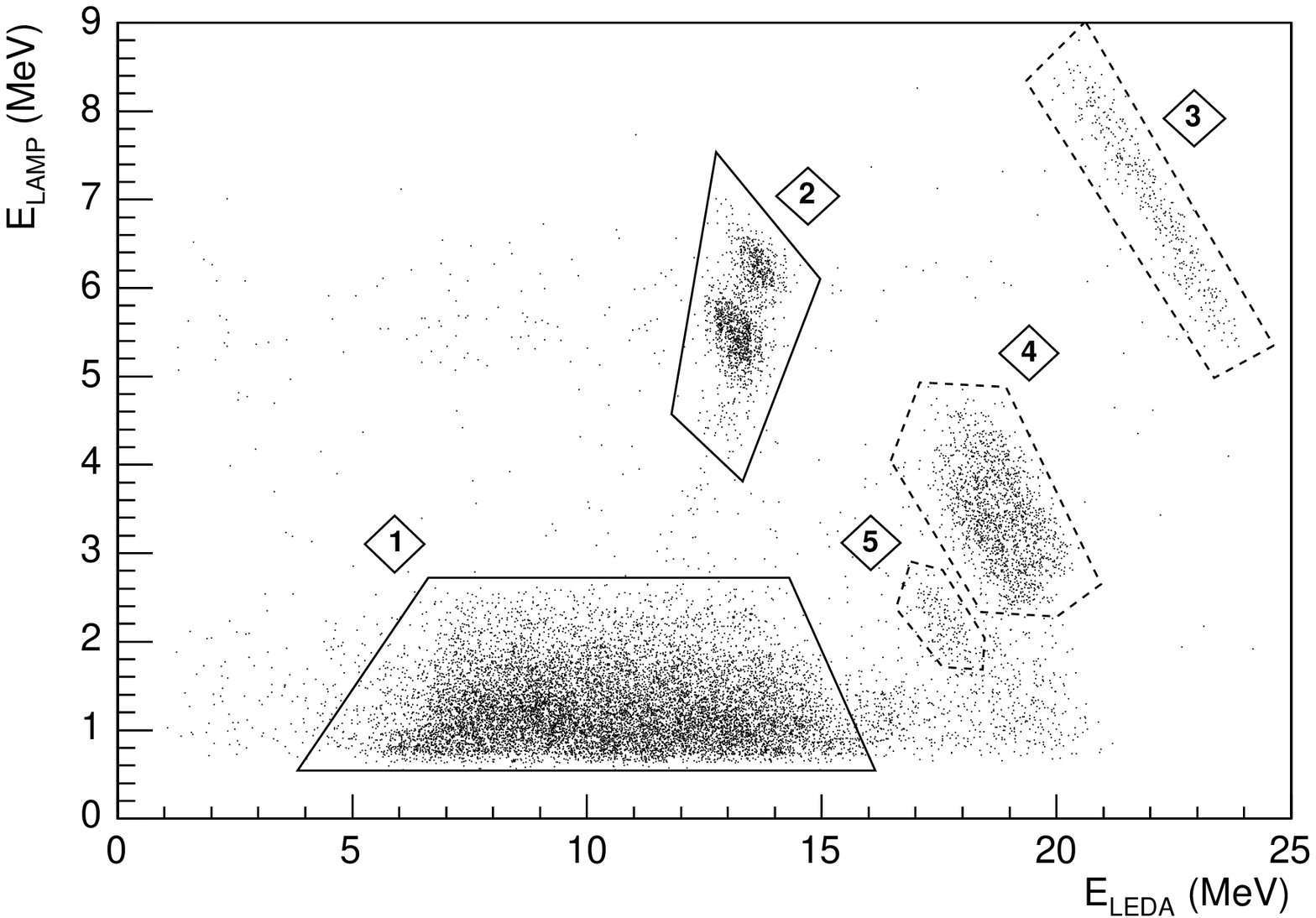}
  \caption{$E_{LAMP} \times E_{LEDA}$ coincidence spectrum showing several
  regions corresponding to the \dpt\ (solid lines regions) and
  d($^{18}$F,$\alpha)^{16}$O (dashed lines regions) reactions. Label 1
  indicates the \dpa\ reaction while label 2 indicates the \dpt\ reaction
  for $^{19}$F excited states below the $\alpha$--threshold. Label 3, 4 and
  5 correspond to different $^{16}$O excited states for the
  d($^{18}$F,$\alpha)^{16}$O reaction.}
  \label{f:coinc}
\end{figure}

Thanks to the low \fo\ energy ($E_{c.m.} = 1.4$~MeV), below the
Coulomb barrier of the \fo\ + $^{12}$C and \fo\ + d systems, only a few
channels are opened: d($^{18}$F,$\alpha)^{16}$O, \dpt, \dpa, d($^{18}$F,n)\nen,
and d($^{18}$F,n$\alpha$)$^{15}$O. Single and coincidence events have
been analyzed but the coincidence condition between LAMP and LEDA (see
Figure~\ref{f:coinc}) allows to distinguish easily the events from
these reactions. The two--body properties of the d($^{18}$F,$\alpha)^{16}$O
reaction can be observed as a linear correlation between the detected energies
in LAMP and LEDA, which is not the case for the three--body \dpa\ reaction.
We performed a further check of the reaction identification by verifying
that the reaction products were (not) detected in the same reaction plane,
indicating a two (three)--body reaction.
                                                                                
\begin{figure}[h]
  \centering
  \includegraphics[width=9cm]{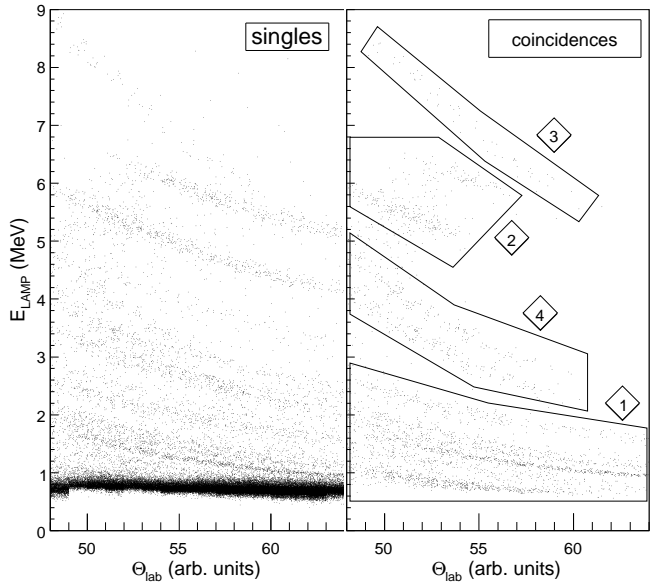}
  \caption{Kinematical bands in LAMP for single and coincidence events.
  Angles are given in arbitrary units. Labels have the same
  meaning as in Fig~\protect\ref{f:coinc}. The different slopes for the
  kinematical bands (region 4 and 1) is the signature of different reactions
  [d($^{18}$F,$\alpha)^{16}$O and \dpt, respectively].}
  \label{f:cinema}
\end{figure}

Kinematical bands (energy versus strip number in LAMP) corresponding to
the events in Fig~\ref{f:coinc} are plotted in Fig~\ref{f:cinema} (right).
For comparison, kinematical bands from single events are also plotted in
Fig~\ref{f:cinema} (left). The same patterns as for coincidence events are
observed but, in addition, since there is no coincidence condition, some 
kinematical bands are completed (region 2) or new (between 2 and 4~MeV).

Protons from the \dpa\ reaction are detected in LAMP in coincidence with
$\alpha$--particle or $^{15}$N in LEDA. Both type of coincidences can be
separated using the time--of--flight information in LEDA, but since the
p--$\alpha$ coincidences represented only 7\% of the total coincidence
events, only the p--$^{15}$N coincidences were analysed.
After selection of the p--$^{15}$N coincidences, the excitation energy of
the decaying \fn\ levels can be kinematically reconstructed from the energies
and angles of the detected protons and the known beam energy. From the
\fn\ excitation energy spectra, and after identifying the populated levels,
the differential cross section in the center--of--mass system was calculated
following the formula:
\begin{equation*}
  \left(\frac{d\sigma}{d\Omega}\right)(\theta) =
  \frac{N_p}{N_{^{18}F} \ N_{d} \ \Delta\Omega}
\end{equation*}
where $N_p$ is the number of detected protons in a given LAMP strip
corresponding to the angle $\theta$, $N_{^{18}F}$ and $N_{d}$ are the
number of incident \fo\ ions and the deuteron content of the target per
unit area, respectively, and $\Delta\Omega$ is the geometrical solid angle.
In the case of coincidences, $\Delta\Omega$ is multiplied by the p--$^{15}$N
coincidence efficiency which is determined by Monte--Carlo simulations
considering an isotropic angular distribution in the $\alpha$--$^{15}$N
center--of--mass system for the $\alpha$--decay of the \fn. However for the
two 3/2$^+$ levels, the $\alpha$--particle is emitted with an $\ell = 1$
angular momentum. Monte--Carlo simulations using $\alpha$--particle angular
distribution corresponding to different sub--magnetic population, gave an
uncertainty related to the normalisation of about 15\%.

\section{Spectra and peak deconvolution\label{sec:Spec}}
\subsection{\fn\ Spectra}
\begin{figure}[h]
  \centering
  \begin{tabular}{c}
    \includegraphics[width=9cm]{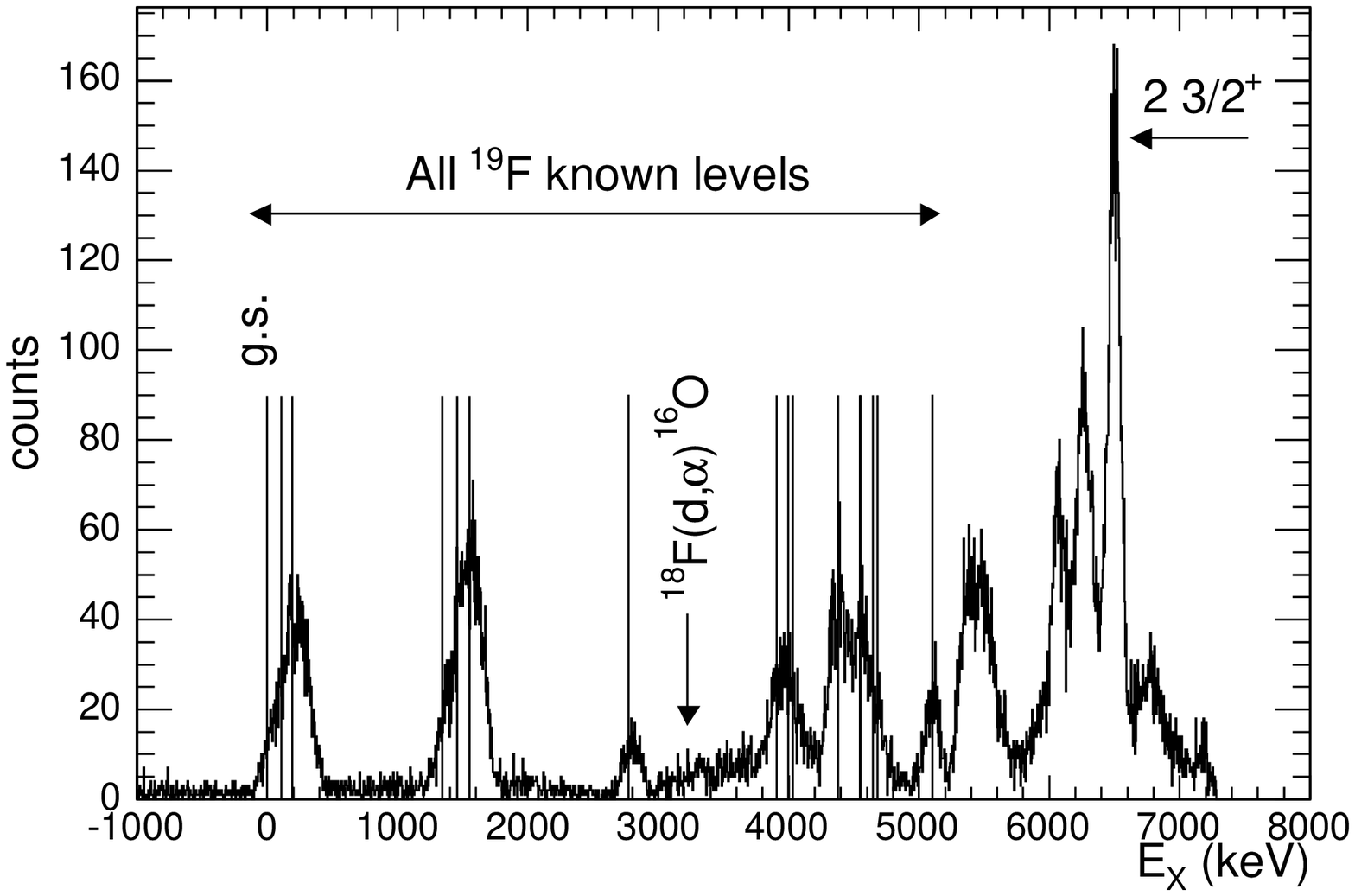} \\
    \includegraphics[width=9cm]{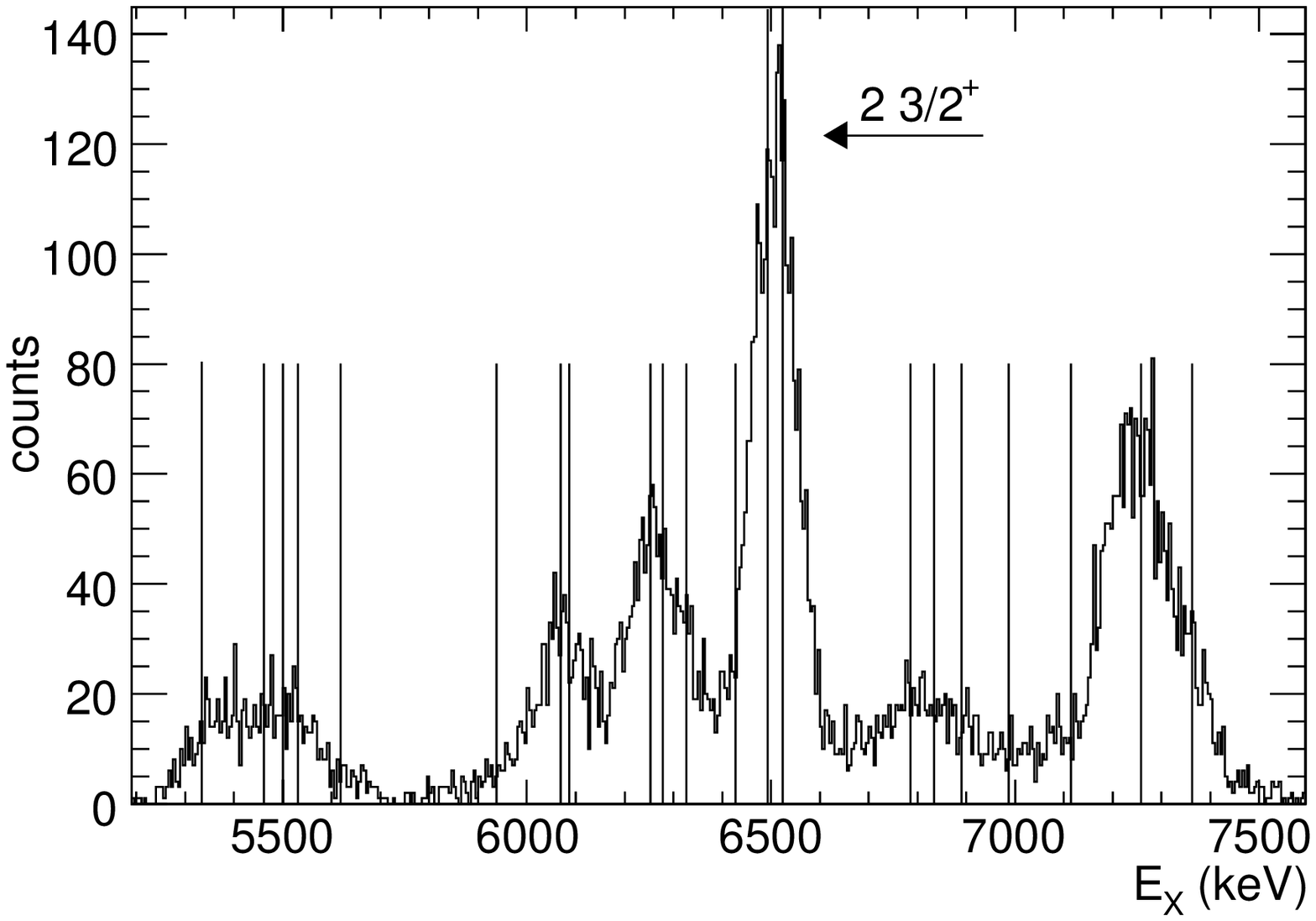} \\
  \end{tabular}
  \caption{{\it Top}: Reconstructed \fn\ spectrum for single events after the
  alpha calibration. Levels
  down to the ground state are populated. Contribution from the
  d($^{18}$F,$\alpha)^{16}$O reaction is shown (see text for details)
  {\it Bottom}: Same for coincidence events, but covering only excitation
  energies above the $\alpha$--particle emission threshold. The doublet of
  3/2$^+$ astrophysical levels around 6.5~MeV is well isolated
  from the other levels. Vertical lines show \fn\ levels (see text).}
  \label{f:spect}
\end{figure}

The reconstructed \fn\ excitation energy spectra are shown in
Figure~\ref{f:spect} for coincidence (bottom) and single (top) events after 
applying the alpha calibration and before the internal calibration (see the
end of the subsection for more details).
Owing to the high coincidence efficiency ($\epsilon = 70\%$), the
coincidence and singles counting statistics were comparable.
Both spectra correspond
to all the events collected in the 6 sectors of the LAMP detector without any
correction for the beam position on target. Because LAMP is very close to the
target, this reconstruction is very sensitive to the beam position on the
target, which is known within $\pm$~2~mm. Hence the angular position of the
strips on a same ring  varies, inducing a different reconstructed excitation
energy for the same detected proton energy. This, in turn, leads to a
deterioration of the resolution when the contribution of the 6 LAMP sectors
are directly summed without correction. To compensate this effect, the 6.5~MeV
peak obtained in coincidence for each LAMP sector is fitted and each spectrum
is shifted relative to the mean position of the 6 sectors.
                                                                                
For the coincidence events, the vertical lines in Figure~\ref{f:spect} (bottom) present the position of the \fn\ levels~\cite{Tilley} populated with low
transferred angular momentum ($\ell \leq 2$). Even if the energy resolution
(FWHM $\approx$ 100~keV) is not sufficient to separate the 20 \fn\ levels that
are represented, the two 3/2$^+$ levels of interest in \fn\ at $E_x = 6.497$
and 6.528~MeV (the analogs of the 3/2$^+$ levels in \nen) are well isolated
from the other groups of levels. Unlike for coincidence events, single events
(Figure~\ref{f:spect} top) are directly selected from a time--of--flight
versus energy spectrum in LAMP, which does not allow to separate proton events
from $\alpha$--particle events coming from the d($^{18}$F,$\alpha)^{16}$O
reaction. For \fn\ excitation energies higher than about 5~MeV, one can observe
the same structure for the single and coincidence spectrum. However, it should
be noted that the 7.2-7.3~MeV peak is less populated than in the coincidence
spectrum because a higher energy threshold for the selection of single events
is used to reduce electronic noise. For lower excitation energy (below the
$\alpha$--threshold), \fn\ levels are observed down to the ground state and
the vertical lines represent all the existing \fn\ levels.

We performed an internal excitation energy calibration using the peaks
corresponding to the well isolated \fn\ levels at $E_x$ = 2.780~MeV, 5.106~MeV
as well as the doublet $E_x$ = 7.262 + 7.363~MeV. In case of the
unresolved peak at 7.3~MeV, its doublet nature was checked by performing
a fit with two components where the position of each component was left as a
free parameter. The experimental energy difference obtained ($\Delta E_x$ =
107~keV) is in very good agreement with the energy difference from the
literature ($\Delta E_x$ = 101~keV) and the hypothetical presence of a third
state in this energy region~\cite{Koz05} was not needed to describe the data.

\subsection{Peak deconvolution\label{sec:Deconvol}}
\begin{figure}[h]
  \centering
  \includegraphics[width=9cm]{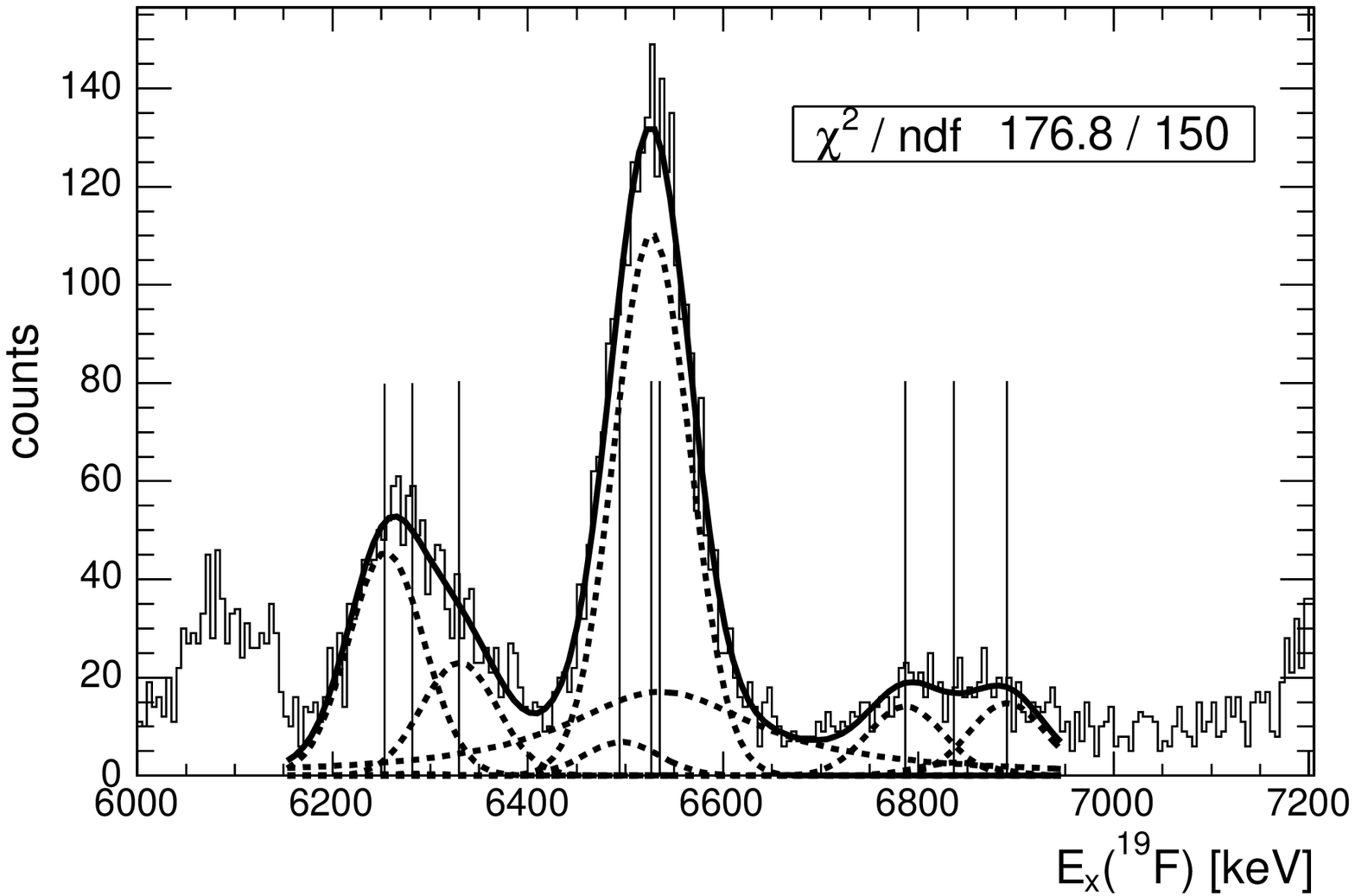} \\
  \caption{Reconstructed \fn\ coincidence spectrum after applying the internal
  energy calibration. The solid line represents the best fit of the spectrum
  and the vertical lines indicate the involved levels. Each level
  contribution is shown in dashed lines. See text for details.}
  \label{f:deconvol}
\end{figure}

The \fn\ excitation energy spectrum obtained after the internal calibration
is shown in Figure~\ref{f:deconvol} as well as a fit to the data between
$E_x = 6150$ and 6950~keV. The fit includes the contribution of two levels
for the 6.5~MeV peak ($E_x = 6497 + 6527$~keV), three levels for the 6.25~MeV
peak ($E_x = 6255 + 6282 + 6330$~keV) and three levels for the 6.9~MeV
peak ($E_x = 6787 + 6838 + 6891$~keV). The free parameters of the fit are
the amplitude of each gaussian and a width common to all the levels. The
level energies are fixed. The excess of counts below the 6.5~MeV peak
is interpreted as the broad 1/2$^-$ level $E_x = 6536$~keV~\cite{Bar05}
($\Gamma = 245$~keV) for which only the intensity is left as a free parameter.
                                                                                
For this broad \fn\ 1/2$^-$ level, we used the values of the energy and
width derived recently~\cite{Bar05} from a reanalysis of previous
data~\cite{Smo61}. Compared to previous adopted values, the width is
similar (245~keV instead of 280~keV) but the energy is about 100~keV
higher ($E_x = 6536$~keV instead of $E_x = 6429$~keV). It should be noted
that the fit obtained with the new parameters is better ($\chi^2$/ndf = 1.19)
than with the previous parameters ($\chi^2$/ndf = 1.54, see Figure~1
in~\cite{Ser04a}). Furthermore, a fit with the same conditions but with
the energy of the broad 1/2$^-$ level as free parameter naturally gives an
energy of $E_x = 6529 \pm 13$~keV, compatible with the new value
from~\cite{Bar05}.

The solid line on Figure~\ref{f:deconvol} shows the best fit and the
dashed lines represent the contribution of each level separately. For the
6.5~MeV peak composed of the two 3/2$^+$ levels, the high-energy level is
favored with a relative contribution of 5\%--95\% for the $E_x = 6497$ and
6527~keV level respectively. A recent experiment using the same reaction
but a higher bombarding energy found an opposite proportion for the
two 3/2$^+$ levels~\cite{Koz05}. In both cases, an internal energy
calibration of the \fn\ excitation energy was used. The differences are
that our energy calibration is an interpolation based on experimental peaks
surrounding the 3/2$^+$ levels of astrophysical interest whereas in
Ref.~\cite{Koz05} the calibration is an extrapolation done for several
laboratory angles, from peaks at lower energies than the levels of
astrophysical interest.
                                                                                
However, since the two levels are not experimentally resolved, different
relative contributions can give similar $\chi^2$ and hence the nominal
relative contribution derived above should be taken with care. Hence we
performed a statistical study aiming at the determination of the upper limit
of the $E_x = 6497$~keV level contribution.
A careful study of the systematic errors (target thickness and detector
dead layer, beam energy, energy loss in the target and in the detector dead
layer, beam centering and detector position) associated to the reconstruction
of the excitation energy, showed that the most important uncertainty is related
to the precise position of the LAMP detector.
Hence three uncertainties were taken
into account: the error associated to the internal energy calibration
($\approx 3$~keV for $E_x \approx 6.5$~MeV), the error associated to the
precise position of LAMP after the internal energy calibration ($\approx
5$~keV for $\pm 1$~mm) and the deconvolution procedure of the 6.5~MeV peak.
The 1--$\sigma$ upper limit for the contribution of the $E_x = 6497$~keV level
for each effect taken separately is 41\%, 34\% and 44\% respectively. When
summed quadratically, we obtain a contribution for the $E_x = 6497$~keV level
of 5$^{+65}_{-5}\%$. It should be noted that there is a 7/2 level of
unknown parity at $E_x = 6554$~keV lying just above the 3/2$^+$ doublet.
We estimated its contribution to be less than 5\%.

From the previous analysis, one cannot exclude that the two 3/2$^+$ levels
are populated in the same proportion. However one expects that these two
levels have different structure properties as it is suggested by an electron 
inelastic scattering measurement on \fn~\cite{Bro85}. Such an experiment
gives access to the tranversal and longitudinal form factors [$F^2(q)$] as a
function of the transfered momentum $q^2$. For such an inelastic scattering,
the form factor depends on the initial and final nuclear states and hence of
the nuclear structure of the observed levels. The longitudinal form factor of
the two 3/2$^+$ levels are different by more than one order of magnitude,
indicating a very different nuclear structure. It is then not surprising that
one of the two levels would be preferentially populated through the \dpt\
transfer reaction, since transfer reactions are very sensitive to the nuclear
structure.

\section{DWBA and compound nucleus components\label{sec:DWBA}}
\subsection{DWBA parameters}
DWBA calculations were performed with the FRESCO code~\cite{FRESCO} in
Finite--Range. For testing the dependency of the results versus the DWBA
parameters, two sets of optical parameters were used: A+A' and B+B' (see
Table~\ref{t:opt}). The optical potentials are written as usual:
\begin{multline*}
  V_{opt} = V_C -V_0 f(r,r_0,a_0) \\ - i \left[ W f(r,r_W,a_W) -
            4 W_D \frac{d}{dr}f(r,r_D,a_D) \right] \\
        + \frac{V_{s.o.}}{r} \left(\frac{\hbar}{m_{\pi}c}\right)
        \vec{\sigma}.\vec{L} \frac{d}{dr}f(r,r_{s.o.},a_{s.o.})
\end{multline*}
where f(r,r$_i$,a$_i$) is a potential well of Wood--Saxon shape with r$_i$
and a$_i$ being the radius and the diffuseness of the interaction
potentials. V$_C$, V$_0$, W, W$_D$ and V$_{s.o}$ are the Coulomb, the volume, 
the volume absorption, the surface absorption and the spin--orbit well depths,
respectively. Usually the optical parameters are determined by fitting
elastic scattering data of the entrance and exit channels. Such data do not
exist for
the d+\fo\ channel, so here the parameters correspond to other nuclear systems
of similar masses and similar energies. The first set of parameters A+A'
comes from a compilation~\cite{Per76}, whereas the second set of parameters
B+B' comes from a similar neutron transfer reaction $^{19}$F(d,p)$^{20}$F
which was studied at the same center--of--mass energy~\cite{Lop64}. In
that study, 11 angular distributions of $^{20}$F excited states
are well described by the optical potential B+B'. For the Finite--Range
calculation, the potential describing the deuteron is of Reid Soft Core
type~\cite{Reid68}. The depth of the volume potential well for the
transferred neutron is automatically adjusted to reproduce its separation
energy in \fn.

\begin{table*}[!hbt]
  \centering
  \caption{Optical potential parameters used for the DWBA analysis. Potentials
  A+A' are from~\protect\cite{Per76} and B+B' are from~\protect\cite{Lop64}.
  Notations are explained in the text.}
  \begin{tabular}{c|ccccccccccccc} \hline \hline
        Optical & r$_C$
        & V$_0$ & r$_0$ & a$_0$
        & W     & r$_W$ & a$_W$
        & W$_D$ & r$_D$ & a$_D$
        & V$_{s.o.} $ & r$_{s.o.}$ & a$_{s.o.}$ \\
        potentials & (fm)
        & (MeV) & (fm) & (fm) & (MeV) & (fm) & (fm)
        & (MeV) & (fm) & (fm) & (MeV) & (fm) & (fm) \\ \hline
        & \multicolumn{13}{c}{\fo\ + d entrance channel} \\
        A & 1.3 & 80.1 & 1.1 & 0.972 & --- & --- & --- & 14.8 & 1.6 & 0.652
          & --- & --- & --- \\
        B & 1.3 & 60.0 & 1.5 & 0.6   & 20  & 1.5 & 0.6 & --- & --- & ---
          & --- & --- & --- \\ \hline
        & \multicolumn{13}{c}{\fn\ + p exit channel} \\
        A' & 1.3 & 47.45 & 1.185 & 0.721 & --- & --- & --- & 7.5 & 0.942 & 0.568           & 5.1 & 1.042 & 0.488 \\
        B' & 1.3 & 50.0 & 1.3 & 0.5 & --- & --- & --- & 8 & 1.3 & 0.5
           & --- & --- & --- \\ \hline \hline
  \end{tabular}
  \label{t:opt}
\end{table*}

It should be noted that even if the energy available in the center--of--mass
system (1.4~MeV) of the \dpa\ reaction is lower that the Coulomb barrier
($E_C = 2.6$~MeV), the emitted protons from the two 3/2$^+$ levels
of astrophysical interest are slightly above the Coulomb barrier of the exit
channel ($E_p = 3.1$~MeV and $E_C = 2.7$~MeV). Hence the term of ``sub--Coulomb
transfer'' does not seem to be justified for the conditions of this reaction.
We have checked this by computing a DWBA calculation in which the nuclear
potentials were switched off, only leaving a Coulomb potential. This
calculation was unable to reproduce the shape of the angular distribution for
the two 3/2$^+$ levels.

\subsection{Compound nucleus component}
Due to the restricted angular range ($\theta_{c.m.} < 50^\circ$), we used
the 9/2$^+$ $E_x = 2780$~keV \fn\ level observed in our data to estimate
the compound nucleus component. Its spin and parity make it difficult to be
populated by a transfer reaction due to the high transferred angular momentum
$\ell = 4$ which corresponds to a 9.2~MeV centrifugal barrier, to be added
to the Coulomb barrier. The DWBA calculation for this level with a maximum
spectroscopic factor of 1 (see Figure~\ref{f:dsig}) fails to describe the
data. Two-step processes such as the excitation of the 3$^+$
state at 937~keV in \fo\ followed by the transfer of a neutron into an
$\ell = 2$ orbital were also computed with the coupled channels code
CCZR~\cite{CCZR}. The contribution of these two-step processes was at most
a few $\mu$b/sr and can not explain the angular distribution of the 9/2$^+$
level. The Hauser--Feshbach
contribution was calculated with the HSFB~\cite{HSFB} code taking into account
the open channels from the compound nucleus $^{20}$Ne decay at our bombarding
energy and normalized to the $E_x = 2780$~keV data. Applying this
normalization to the 5/2$^+$ $E_x = 5106$~keV \fn\ level describes successfuly
the data which does not seem to be described by a dominant direct reaction
mechanism (see Figure~\ref{f:dsig}).

\section{Discussion\label{sec:Results}}
\begin{figure}[!ht]
  \centering
  \begin{tabular}{cc}
     \includegraphics[width=7cm]{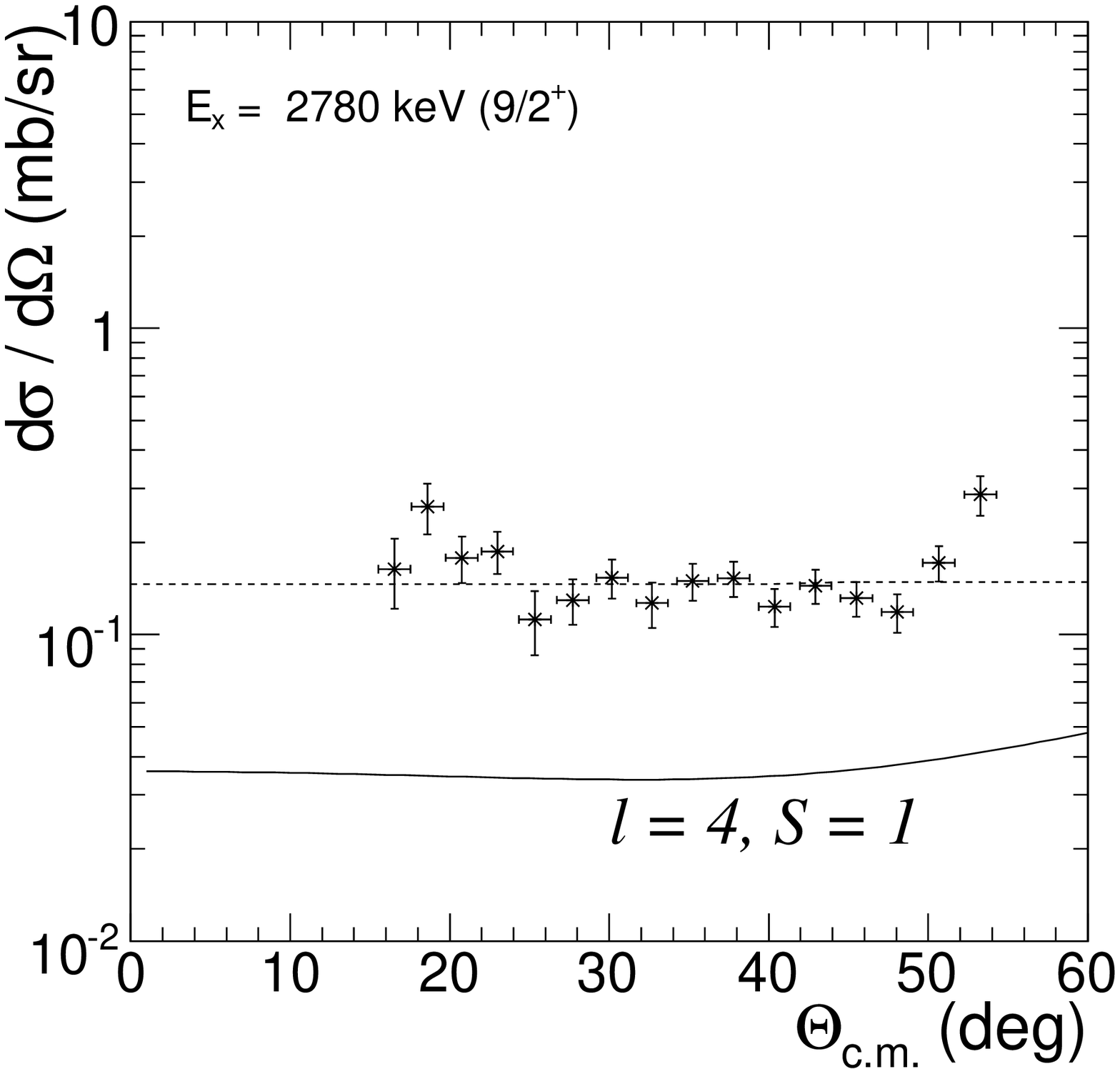} &
     \includegraphics[width=7cm]{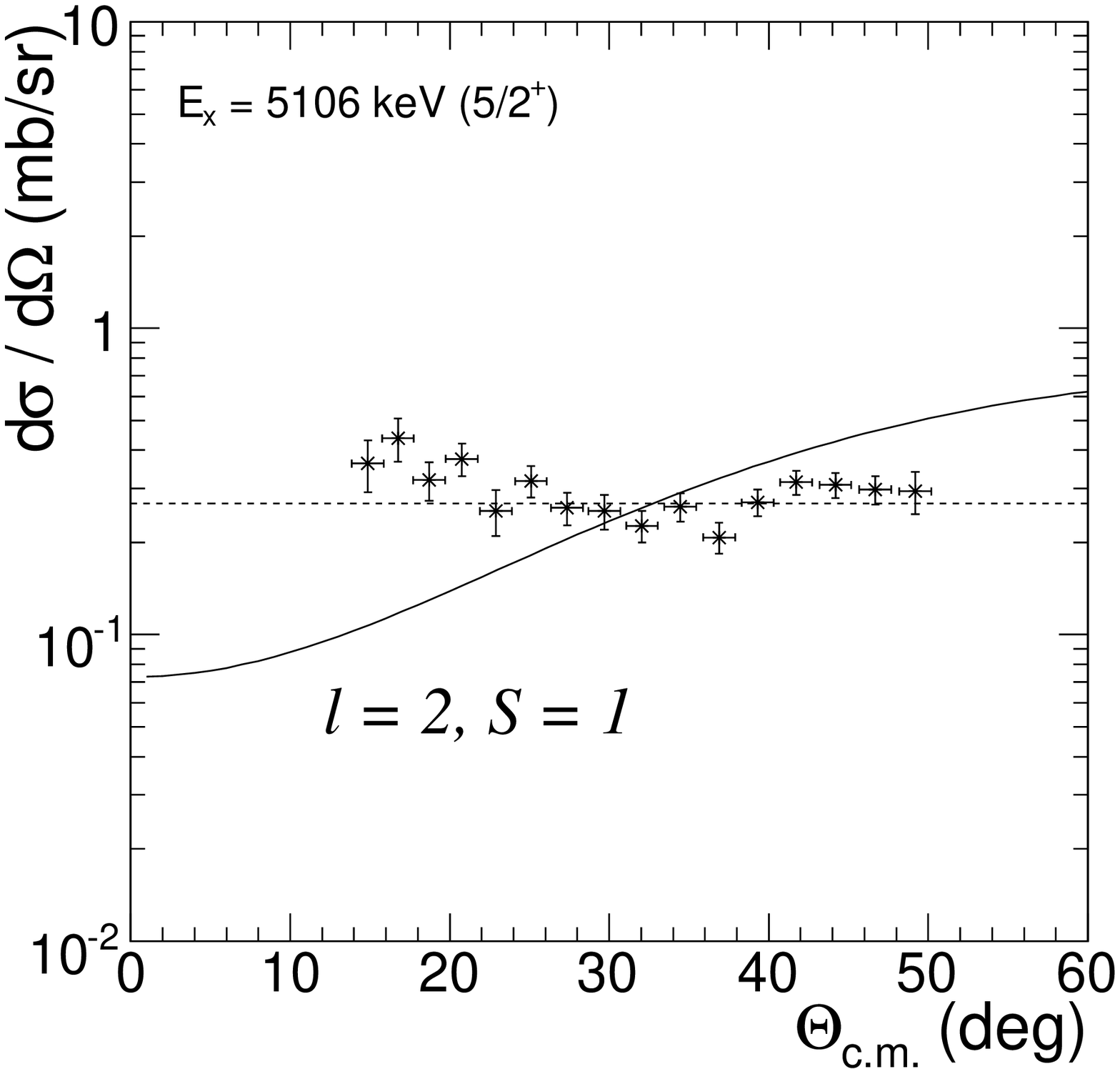} \\
     \includegraphics[width=7cm]{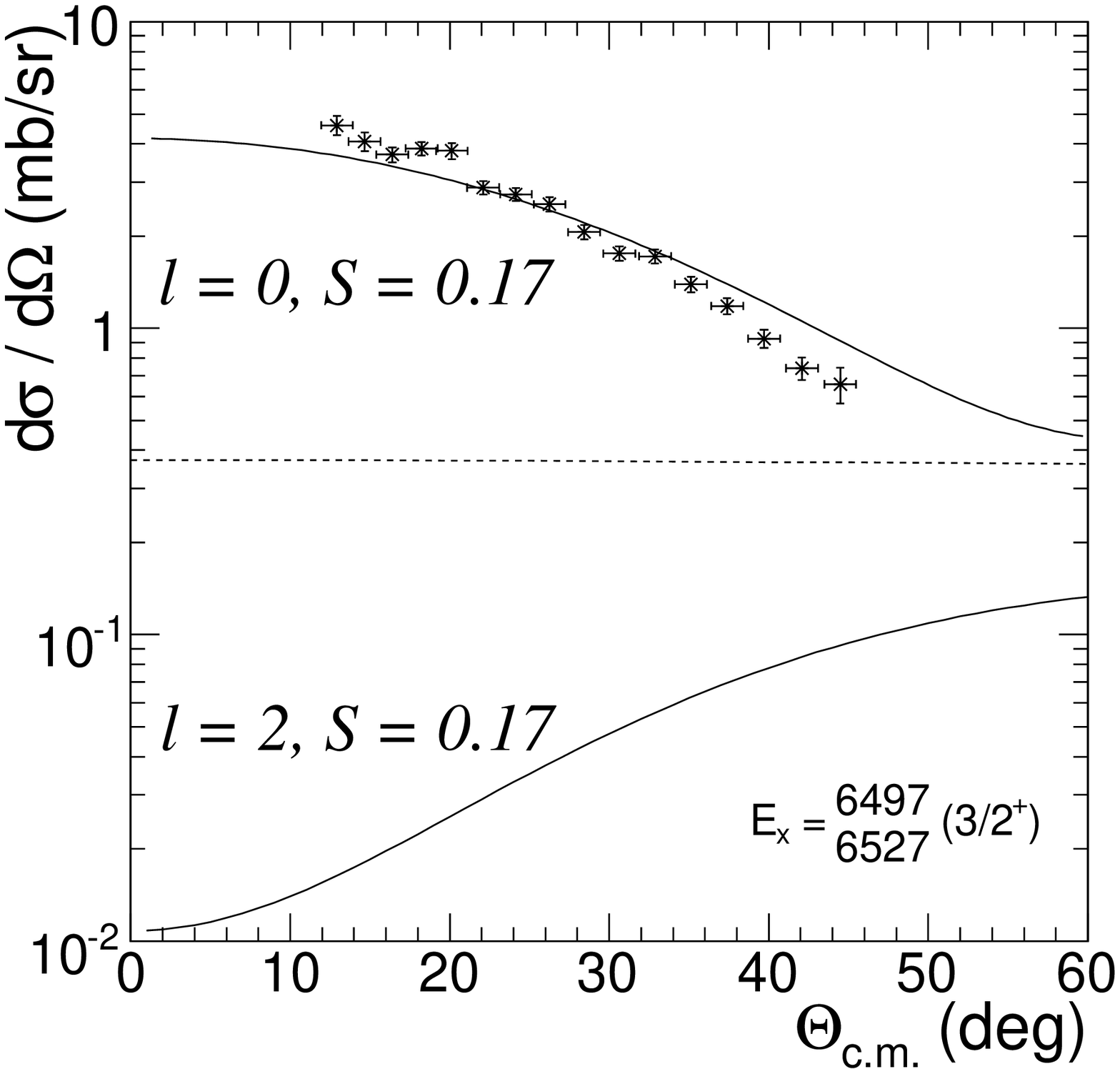} &
     \includegraphics[width=7cm]{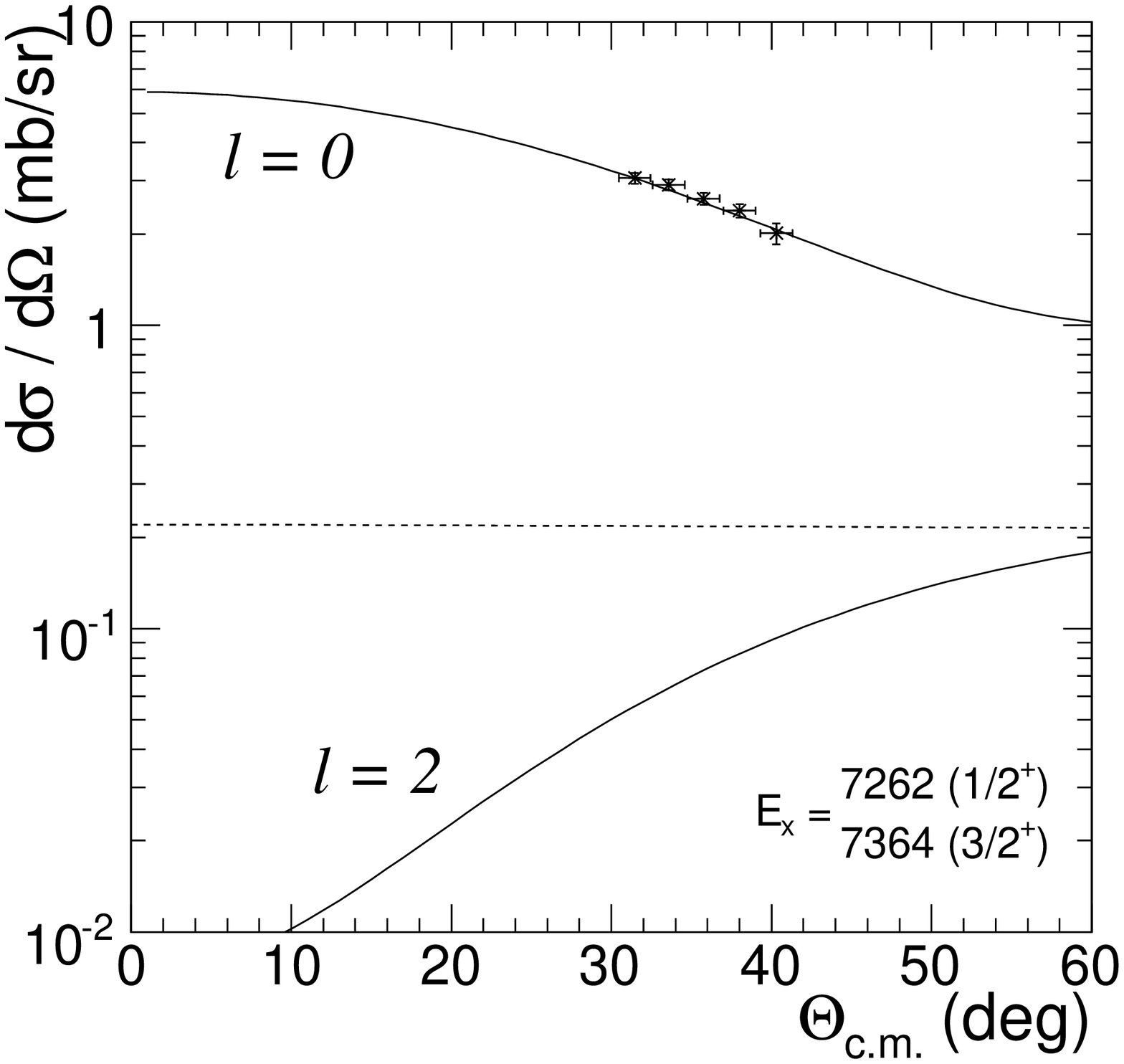} \\
  \end{tabular}
  \caption{Angular distributions for the \fn\ levels at 2.780, 5.106~MeV,
  the doublet at 6.497 and 6.537~MeV, and the levels at 7.262 and 7.364~MeV,
  populated by the \dpt\ transfer reaction.
  Dashed lines represent the compound nucleus component and solid line are the
  results of the direct contribution calculated using a Finite-Range DWBA
  analysis and the B+B' optical potentials.}
  \label{f:dsig}
\end{figure}

The angular distributions of single data for the $E_x(\fn) = 2780$ and 5106~keV
levels and of coincidence data for the doublet $E_x(\fn) = 6497$ and 6527~keV
and the two $E_x(\fn) = 7262$ and 7364~keV levels are shown in
Figure~\ref{f:dsig}. While the angular distributions for the two levels at
$E_x = 2780$ and 5106~keV are quite flat, supporting a compound nucleus type
reaction mechanism, the angular distribution of the 3/2$^+$ doublet shows a
strong variation of about one order of magnitude in a small center--of--mass
angular range, which is typical of a direct reaction mechanism. The
spectroscopic factors obtained from our analysis are displayed in
Table~\ref{t:c2s}.
                                                                                
\begin{table}[h]
  \centering
  \caption{Spectroscopic factors for the \fn\ levels analyzed in DWBA.}
  \begin{tabular}{cccc}  \hline \hline
      $E_x$(\fn)  [MeV]  &  J$^{\pi}$  &  $\ell_n$  &  $C^2S$  \\ \hline
         $\left. \begin{array}{c} 6.497 \\ 6.527 \end{array} \right\}$
             & 3/2$^+$ & 0 & 0.17 \\
         6.536    &   1/2$^-$   & 1 & $<$ 0.33 \\
         7.262    &   3/2$^+$   & 0 &   0.15   \\
         7.363    &   1/2$^+$   & 0 &   0.22   \\ \hline \hline
   \end{tabular}
  \label{t:c2s}
\end{table}

\subsection{Levels at $E_x$ = 6.497 and 6.527~MeV in \fn}
The two 3/2$^+$ levels can be populated by both $\ell = 0$ and $\ell = 2$
transfers, however the fit of the DWBA calculations to the data indicates
that only the $\ell = 0$ component is contributing (see Figure~\ref{f:dsig})
at such low center of mass energy. At much higher incident energy, a
$\ell = 2$ component was requested to reproduce transfer data~\cite{Koz05}.
Taking into account the compound nucleus contribution as explained above,
we obtain for $\ell = 0$ a spectroscopic factor of $S(6.497+6.527) = 0.17$
(to be compared with $S = 0.12$ from a recent experiment~\cite{Koz05}). This
value supersedes our previous value of $S(6.497+6.527) = 0.21$~\cite{Ser03a}:
the difference comes from the new Finite--Range DWBA calculations and the
inclusion of the Hauser--Feshbach contribution. The fact that the two sets of
optical parameters A+A' and B+B' give very similar angular distributions and
spectroscopic factors makes us to feel confident about the present results.
The analysis of the single data events for this 3/2$^+$ doublet allowed us to
derive the corresponding angular distribution and to compare it with the
coincidence one. They agree well, both in the shape and in the normalisation,
hence giving strong confidence in the coincidence solid angle determined by
Monte--Carlo simulations.
                                                                                
The uncertainties associated to the spectroscopic factors have been evaluated
to be of 25\%. The major source of uncertainty comes from the DWBA method
(20\%) and is related to the choice of the optical parameters in the entrance
channel which cannot be constrained by experimental data. Another source of
uncertainty (15\%), which has already been discussed, arises from the angular
distribution of the emitted $\alpha$--particle in the \fn\ center--of--mass
frame. The uncertainty on the normalization contributes to 7\%
(Section~\ref{sec:Spec}).

\subsection{Level at $E_x$ = 6.536~MeV in \fn}
Due to its large total width $\Gamma = 245$~keV~\cite{Bar05}, this 1/2$^-$
level can play an important role in the \pa\ reaction since it covers the
whole Gamow peak at novae temperatures. But its large width also makes it
difficult to be observed. However, making the hypothesis that all the excess
of counts below the 6.5~MeV peak comes from the 1/2$^-$ level, an upper
limit to its contribution could be derived. This hypothesis does not seem
unrealistic because no counts are observed in the coincidence spectrum in the
$E_x = 5.8$~MeV region (see Figure~\ref{f:spect} bottom), where no \fn\ levels
exist. Hence this seems to indicate that the excess of counts below the
6.5~MeV peak has a physical origin.
                                                                                
The number of counts corresponding to the population of the 1/2$^-$ level
is obtained from the global fit presented in Figure~\ref{f:deconvol} where
this level is described by a lorentzian shape whose amplitude is a free
parameter. Then, assuming a $\ell = 1$ transfer for populating this level,
one obtains an upper limit for the spectroscopic factor of $S(6.536) < 0.33$.
The difference with respect to our previous value ($S(6.536) <
0.15$~\cite{Ser03a}) comes from the new position and width of this
level~\cite{Bar05} as well as from the DWBA calculation, now made in
Finite--Range.

\subsection{Levels at $E_x$ = 7.262 and 7.364~MeV in \fn}
The experimentally unresolved \fn\ levels at 7.262 and 7.364~MeV have
a 3/2$^+$ and 1/2$^+$ spin and parity, respectively, and both can be
populated by a $\ell = 0$ or $\ell = 2$ neutron transfer. However the
DWBA calculations fitted to the data showed that only the $\ell = 0$
is contributing (see Figure~\ref{f:dsig}). These two levels have the
highest excitation energy which can be achieved with the beam energy
used here, and correspond to the less energetic protons detected ($E_p
< 1$~MeV). Part of the kinematical band of these two levels, which are
not experimentally resolved, is cut by the electronic threshold. Hence
only the five outer strips of LAMP were considered in the analysis. The
extracted angular distribution does not allow to deduce the relative
contribution of the two levels since the differential cross section
has the same shape due to the same $\ell = 0$ transfer but a different
normalisation (a factor of 2) due to the spin factor. However a two--peak
fitting analysis of the 7.3~MeV excitation energy peak indicates a relative
contribution of 57\%--43\% for the 7.262 and 7.364~MeV levels, allowing to
obtain a set of spectroscopic factors $S(7.262) = 0.15$ and $S(7.364) = 0.22$.
The compound nucleus component which has been taken into account has a
negligible cross section ($\approx 0.1$~mb/sr).

\subsection{Analog level of the $E_x(\nen)$ = 7.076~MeV}
Up to now, the analog level in \fn\ of the \pa\ resonance at $E_r = 665$~keV
[$E_x(\nen) = 7.076$~MeV, J$^\pi$ = 3/2$^+$] is still not identified. A
gamma--ray measurement~\cite{But98} via the $^{15}$N($\alpha$,$\gamma$)\fn\
reaction found a level at 7.101~MeV in \fn\ with a width $\Gamma_\alpha = 28$
~keV. The spin and parity was deduced to be 3/2$^+$ by observing gamma--ray
transitions to a 3/2$^-$ level. Assuming equality of reduced
$\alpha$--widths between analog states, Butt~et~al.~\cite{But98} find
$\Gamma_{\alpha} \approx 30$~keV for the \nen\ level at $E_x = 7.076$~MeV,
which is in good agreement with the value of $\Gamma_{\alpha} = 24$~keV
from~\cite{Bar01a}.
                                                                                
However, the assignation of a 3/2$^+$ spin and parity to the $E_x(\fn) =
7.101$~MeV level has been questioned by Fortune and Sherr~\cite{For00} who
pointed out that this level could be a doublet of spins 1/2--5/2 or 1/2--7/2.
Furthermore, these authors predict the position of the analog level at higher
energy around $E_x = 7.4 \pm 0.1$~MeV.
                                                                                
Assuming here that the 7.101~MeV level in \fn\ is the analog of the
$E_x(\nen) = 7.076$~MeV and assuming equality between dimensionless reduced
widths [$\Theta^2_n(\fn) = \Theta^2_p(\nen)$], we obtain $\Theta^2_n(\fn)
= 0.14$. This value is comparable to the one from the two 3/2$^+$ levels
of astrophysical interest ($\Theta^2_n = 0.1$) and should hence appear in
our spectra as a peak of similar intensity as the 6.5~MeV peak, but our data
(Figure~\ref{f:spect}, bottom) do not show any evidence of a peak around
7.1~MeV. Furthermore it has been shown recently that the spin of this
resonance was unlikely to be 3/2$^+$~\cite{Bar05}.
                                                                                
The 7.3~MeV group in \fn\ is composed of two states at 7.262~MeV (3/2$^+$)
and 7.368~MeV (1/2$^+$) whose spectroscopic factors are determined here to
be $S = 0.15$ and 0.22, respectively (see Table~\ref{t:c2s}). The spectroscopic
factor value for the 3/2$^+$ state is very close to the one of the level at
7.076~MeV in \nen, and seems then to indicate that the $E_x(\fn) = 7.262$~MeV
is a good candidate for the analog level of the 7.076~MeV in \nen. However, it
should be noted that the total width of the 7.262~MeV level in \fn\ calculated
from the $E_x(\nen) =
7.076$~MeV is $\Gamma = \Gamma_{\alpha} \approx 28$~keV which seems
incompatible with the measured value $\Gamma < 6$~keV~\cite{Tilley}.

\subsection{Proton widths in \nen\label{sec:gammap}}
The obtained neutron spectroscopic factors are for \fn\ levels whereas the
proton widths of astrophysical interest correspond to their analog levels
in \nen. Following Utku et al.~\cite{Utk98}, we assume that the two
3/2$^+$ levels $E_x =6497$ and 6527~keV and the 1/2$^-$ level $E_x = 6536$~keV
populated in \fn\ are the analog states of the 3/2$^+$ levels $E_x(\nen) =
6419$ and 6449~keV, and of the $E_x(\nen) = 6437$~keV, respectively, even if
the spins and parity of \nen\ states has not been determined experimentaly.
Unlike other analog levels in \fn/\nen\, where the assignation comes from the
fact that the levels are populated with comparable statistics by mirror
reactions, the case of the two 3/2$^+$ and 1/2$^-$ levels only relies
on the similarity of their energy position and of their total width,
respectively. The two 3/2$^+$ levels are both separated by 30~keV in \fn\
and \nen\ at similar excitation energies, whereas for the 1/2$^-$ level
$\Gamma_\alpha(\nen) = 220~\mathrm{keV} \approx \Gamma_\alpha(\fn) =
245~\mathrm{keV}$.
                                                                                
\begin{figure}[h]
  \centering
  \includegraphics[width=9cm]{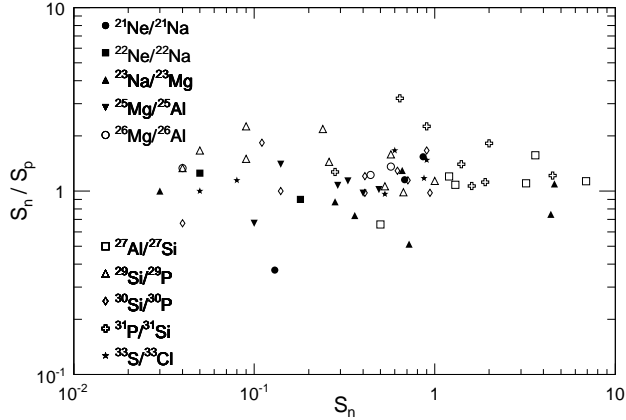} \\
  \caption{Comparison of neutron and proton spectroscopic factors for analog
  levels of the $sd$ shell nuclei. Data comes from~\protect\cite{Lan60}.}
  \label{f:c2s}
\end{figure}

It is of common use in nuclear astrophysics to assume equality of spectroscopic
factors between analog states. Hence for the 3/2$^+$ states in \nen, we
obtain a proton spectroscopic factor $S = 0.17$ and for the 1/2$^-$ state, we
obtain an upper limit $S < 0.33$. However, the uncertainty associated to this
practice is estimated to be a factor of two from the comparison of the values
of neutron and proton spectroscopic factors for all the analog states of the
$sd$ shell nuclei (Figure~\ref{f:c2s}).
                                                                                
The proton widths are calculated using the following relations:
\begin{equation*}
   \Gamma_p = C^2S \ \Gamma_{s.p.} \ \ \ \mathrm{with} \ \ \
   \Gamma_{s.p.} = \frac{\hbar^2 s}{\mu} |\mathcal{R}_{s.p.}(s)|^2 P_l,
\end{equation*}
where $\Gamma_{s.p.}$ is the single--particle width, $s$ is the interaction
radius ($s = 4.5$~fm) and $\mathcal{R}_{s.p.}$ is the proton radial wave
function in the p+\fo\ system calculated with classical parameters for
the Wood--Saxons nuclear potential: ($r$, $a$)~=~(1.25, 0.65)~fm, where $r$
and $a$ are the radius and diffuseness of the potential. As compared
to direct determination of the proton width, a statistical study of $sd$
shell nuclei showed an agreement between direct and indirect method of
about 30\%~\cite{Hal04}. Using the above prescription we obtain for the
3/2$^+$ doublet in \nen, $\Gamma_p(6.419) = 3.4\times10^{-37}$~keV and 
$\Gamma_p(6.449) = 1.9\times10^{-37}$~keV assuming that the total strength
of the doublet is put on each level respectively. As far as the levels above 
$E_x$ = 7~MeV are concerned and if we assume that the levels $E_x(\fn) = 
7.262$~MeV and $E_x(\nen) = 7.238$~MeV are analogs~\cite{Utk98} we obtain
$\Gamma_p(7.238) = 10.3$~keV which is larger than the adopted upper limit
($\Gamma_p < 4$~keV~\cite{Utk98}). For the analog state of the $E_x(\fn) =
7.364$~MeV 1/2$^+$ level, no 1/2$^+$ levels in \nen\ have been observed so far
at an excitation energy around 7.3~MeV.

\section{CONCLUSIONS: REMAINING UNCERTAINTIES\label{sec:Sfac}}
Our analysis showed that the two expected low--lying 8 keV and 38 keV
proton resonances ($E_x = 6419$ and 6449~keV in \nen) must be considered for
the calculation of the \pa\ reaction. Here we focus on the remaining
uncertainties before firm reaction rates can be computed for the \pa\ and \pg\
reactions.
 
Due to the small energy difference (30~keV) between the two 3/2$^+$
levels in \fn\ ($E_x = 6497$ and 6527~keV) and the two resonances in \nen\
($E_r = 8$ and 38~keV), the assignment of analog pairs is not straigthforward
and one cannot discard the possibility of an inversion. Hence the 
identification of the main contributing \fn\ level is not so crucial for the 
determination of the associated proton widths in \nen. So two cases will have 
to be considered, the whole strength at the $E_r$ = 8~keV resonance or at the 
$E_r$ = 38~keV resonance.
                                                                                
The second uncertainty arise from interference effects between the two $E_r$ =
8 and 38~keV resonances and the 3/2$^+$ resonance at $E_r = 665$~keV
(corresponding to the \nen\ level $E_x = 7.076$~MeV). The sign of these
interferences is totally unknown and for the destructive case, the
S--factor would become extremely small between 100 and 300~keV, which
corresponds to Gamow peak energies for temperatures achieved in novae ($50 <
T_6 < 350$). As usual, the notation $T_x = y$ corresponds to the temperature
$y \times 10^{x}$~K. 
                                                                                
Another uncertainty on the reaction rates comes from the poor knowledge of the
\nen\ alpha widths, $\Gamma_\alpha$, for the $E_r = 8$ and 38~keV resonances.
Since there are no measurements of these $\Gamma_\alpha$ in \nen, they are
usually deduced from the analog states in \fn. However differences as large
as a factor 10 have been observed for alpha widths of analog levels for the
\fn\ / \nen\ mirror nuclei~\cite{Oli97}. Calculations of constructive and
destructive interferences for different values of alpha widths showed a strong
impact on the \pa\ reaction rate~\cite{Ser03b}.

Finally the energy of the two $E_r = 8$ and 38~keV resonances in \nen\ is 
not known better than 6~keV~\cite{Utk98}. This effect is far less important 
than the one of the unknown alpha widths because the position of the 
resonances are below the energies corresponding to the Gamow peak for 
temperatures relevant to \fo\ production in novae.
                                                                                
In conclusion we reported here on our latest results concerning the 
one--nucleon transfer reaction \dpa\ used to study \fn\ analog states of 
the astrophysically important states in \nen. A DWBA analysis of the data 
showed that the two 3/2$^+$ low--energy resonances in \nen, $E_r = 8$ and 
38~keV have a relatively large spectroscopic factor ($S$ = 0.17) and hence 
cannot be neglected in calculating the \pa\ and \pg\ reaction rates. We made 
a careful analysis of the impact on the reaction rates of remaining 
uncertainties. This includes interferences between the low--lying states 
and the 3/2$^+$ $E_r = 665$~keV resonance, the unknown $\alpha$--widths in 
\nen\ and the possibility of inversion for the analog levels between \fn\ 
and \nen. A new challenging experiment to investigate the \pa\ reaction at 
energies below 0.6~MeV in the center--of--mass will be soon performed at 
the Louvain-la-Neuve CYCLONE facility.

\section*{Acknowledgments}
This work has been supported by the European Community-Access to Research
Infrastructure action of the Improving Human Potential Program, contract
N° HPRI-CT-1999-00110, and the Belgian Inter--University Attraction Poles
P05/07. One of us (P.L.) is a Research Director of the National Fund for
Scientific Research, Brussels. We wish to thank Jean-Pierre Thibaud for
very fruitful discussions concerning the data analysis and his careful
reading of the manuscript. We also thank Jordi Jos\'e and Margarita
Hernanz for very interesting discussions concerning the astrophysical
implications of the present work.

\bibliographystyle{npa}
\bibliography{nuclear,astro}

\end{document}